# Abruptly autofocusing waves enter space-time

Nikolaos K. Efremidis[1,2,*] and Demetrios N. Christodoulides[3]

[1]Department of Mathematics and Applied Mathematics, University of Crete, Heraklion, Crete 70013, Greece

[2]Institute of Applied and Computational Mathematics, FORTH, Heraklion, Crete 70013, Greece

[3]Ming Hsieh Department of Electrical and Computer Engineering, University of Southern California, Los Angeles, California 90089, USA

*Corresponding author: nefrem@uoc.gr

Focusing is one of the most fundamental operations in optics: it determines how precisely optical energy can be delivered, how strongly light can interact with matter, and how sharply structures can be imaged, modified or manipulated. The range of applications is vast, extending from optical microscopy, lithography and laser micromachining to multiphoton excitation, optical trapping and nonlinear light–matter interactions. The simplest and most familiar route to optical focusing is to pass a Gaussian beam through a lens, thereby imposing a quadratic phase that gradually concentrates the optical field near the focal plane [1].

In 2010, a new focusing principle was introduced based on the inward collapse of a caustic toward the optical axis [2]: as the field propagates, its high-intensity ring, or shell in the spherical case, contracts radially and eventually concentrates its power at a prescribed focus, as illustrated in the left panel of Fig. 1. These "abruptly autofocusing waves" were originally formulated using Airy profiles arranged in circular and spherical geometries. Importantly, compared with the smooth Lorentzian-like profile of a Gaussian focus, the abruptly autofocusing beam maintains a low peak intensity over most of the propagation and then exhibits a sudden, high-contrast intensity surge near the focus.

Two different geometries of abruptly autofocusing waves were introduced in the original theory [2]. These constructions are built on Airy profiles, whose planar finite-energy form are shape-preserving accelerating beams [3]. The first is the circular two-dimensional Airy wave that depends on the transverse circular coordinate $\rho = (x^2 + y^2)^{1/2}$ and propagates along the $z$-direction. The second is the experimentally more challenging three-dimensional abruptly autofocusing wave with spherical symmetry in space-time where $r = (x^2 + y^2 + (\gamma t)^2)^{1/2}$ is the radial coordinate and the beam propagates along the $z$-coordinate. Interestingly, spherical abruptly autofocusing waves admit an exact closed-form solution in terms of Airy functions, a property that is absent from their circular counterparts. Abruptly autofocusing waves are not limited to Airy-type excitations; they can also be generated from engineered caustic profiles with radial symmetry, offering greater flexibility in controlling the beam trajectory and focusing contrast [4].

The generation of circular abruptly autofocusing beams is conceptually straightforward: the main requirement is to imprint an appropriate phase mask so as to synthesize the desired beam profile either in the Fourier plane [5] or directly in real space. One year after the theoretical prediction, two experimental studies reported the realization of radially symmetric two-dimensional abruptly autofocusing waves. In particular, in [6] abruptly autofocusing waves were observed and used to produce localized ablation spots on the rear surface of thick samples, illustrating their potential for energy delivery inside or through extended media. Shortly afterwards, a second experiment demonstrated the use of abruptly autofocusing beams for optical manipulation, guiding and transporting microparticles along the primary rings of a radially symmetric Airy beam [7]. More generally, the ability of abruptly autofocusing beams to produce strong localization at a prescribed distance has enabled a variety of applications in particle manipulation, material modification and microfabrication, THz generation, and nonlinear processes [8–11], see also the review article [12].

The spherical counterpart, however, remained experimentally elusive for more than fifteen years. Its recent realization by Cao et al. brings abruptly autofocusing waves into the full space-time domain [13]. This transition is by no means trivial. Two-dimensional spatial encoding $F(x, y)$ is technically accessible, as it can often be achieved with a phase element acting in the transverse plane. Space-time light sheets are also experimentally feasible: by dispersing the pulse spectrum across a spatial light modulator, one can correlate one transverse spatial coordinate with frequency and synthesize wavepackets of the form $F(x, t)G(y)$ [14]. Generating a three-dimensional space-time wavepacket $F(x, y, t)$ with a non-separable field structure is significantly more complex, since both transverse spatial dimensions must be coupled to the temporal degree of freedom. The recent experiment by Cao et al. demonstrates an elegant solution to this problem [13]. Rather than attempting to synthesize the spherical Airy wavepacket voxel by voxel, they used the radial symmetry of the optical field and constructed it through a sequence of lower-dimensional transformations. The spherical Airy wavepacket is not an arbitrary function of $x$, $y$, and $t$, but depends on the combined space-time radius ($r = (x^2 + y^2 + (\gamma t)^2)^{1/2}$). The experimental approach uses a two-dimensional space-time Airy precursor, generated by applying a spatiotemporal hologram in a spatial–spectral plane. The field is then extended along the second transverse dimension and subjected to a log-polar-type optical transformation, which folds the space-time sheet into the desired spherical geometry [see Fig. 1]. The experiment combines free-space diffraction with an SLM-imposed spectral phase that emulates the anomalous dispersion required for ideal spherical Airy propagation. The dynamics reveal a tightly focused beam in three-dimensional space-time with contrast significantly larger than that of the corresponding equal-envelope Gaussian beam.

Several challenges remain. The present implementation relies on pixelated spatial light modulators, finite apertures, limited spectral bandwidth and approximate coordinate transformations. These constraints limit the achievable focal size, energy efficiency and accessible parameter range. The log-polar mapping also introduces Jacobian amplitude factors and residual phases that must be compensated carefully. Future implementations using custom diffractive optics, metasurfaces, integrated photonic elements or adaptive multi-plane light conversion could improve the fidelity and efficiency of the transformation. Such developments may allow spherical Airy wavepackets with tighter foci, higher powers, tailored polarization or orbital angular momentum, and enhanced resilience in random media.

Three-dimensional variants of abruptly autofocusing waves —such as spherical Airy wavepackets or circular autofocusing beams combined with temporal focusing schemes— could provide higher focusing contrast, tighter localization, and improved robustness in complex media. These features are especially relevant to medical optics, microscopy, particle manipulation, microfabrication, and nonlinear processes, where one seeks to localize optical energy at a prescribed position while minimizing its impact elsewhere.

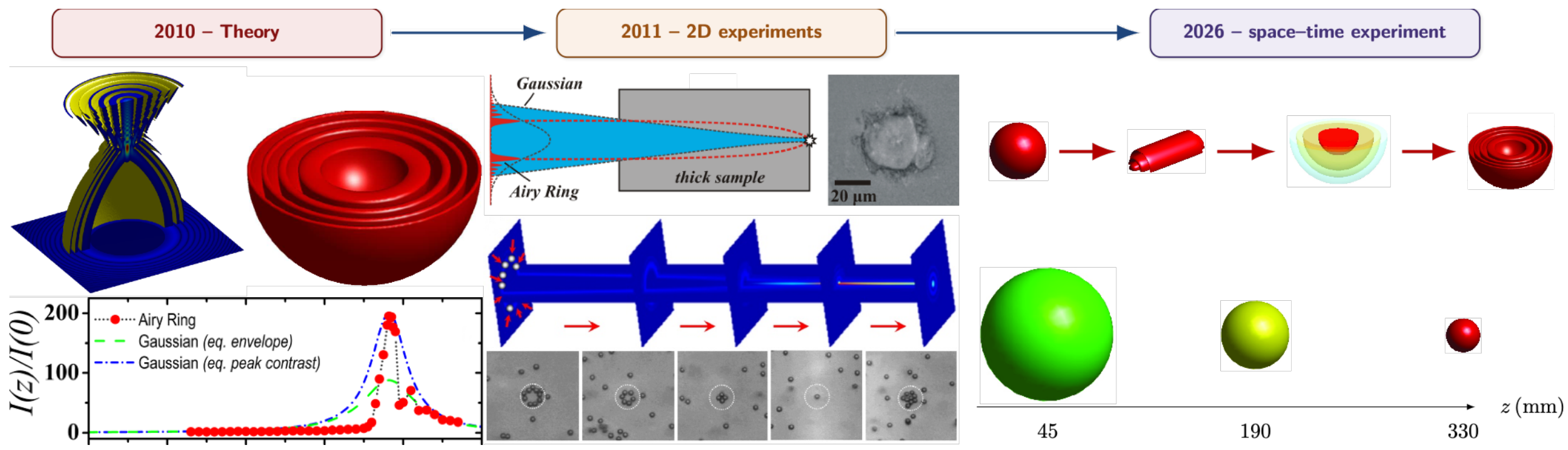

**Figure 1:** Timeline of abruptly autofocusing waves. In 2010 circular and spherical abruptly autofocusing waves were theoretically predicted [2]. These optical waves maintain a low peak intensity during most of their propagation, before increasing rapidly by orders of magnitude near the focus. In 2011, circular abruptly autofocusing waves were experimentally observed and used to create localized ablation spots [6], as well as to transport and manipulate microparticles [7]. Recent advances in spatiotemporal wavepacket synthesis have now enabled the experimental realization of the spherical counterpart, bringing abruptly autofocusing waves into the full space-time domain [13]. Panels adapted from Refs. [2,6,7].